\documentclass[conference]{IEEEtran}
\IEEEoverridecommandlockouts
\usepackage{cite}
\usepackage{amsmath,amssymb,amsfonts}
\usepackage{algorithmic}
\usepackage{graphicx}
\usepackage{textcomp}
\usepackage{xcolor}
\usepackage{longtable}
\usepackage{hyperref}
\usepackage{array}
\usepackage{booktabs}
\def\BibTeX{{\rm B\kern-.05em{\sc i\kern-.025em b}\kern-.08em
    T\kern-.1667em\lower.7ex\hbox{E}\kern-.125emX}}
\begin{document}

\title{Balancing hydrogen delivery in national energy systems: impact of the temporal flexibility of hydrogen delivery on export prices\\
\thanks{This work has been submitted to the IEEE for possible publication. Copyright may be transferred without notice, after which this version may no longer be accessible. These results are based on work done for the HYPAT project, funded by the German Federal Ministry of Education and Research (03SF0620A).}
}

\author{
\IEEEauthorblockN{Hazem Abdel-Khalek\textsuperscript{1,2,3,a}, Eddy Jalbout\textsuperscript{2,4}, Caspar Schauß\textsuperscript{5}, Benjamin Pfluger\textsuperscript{2}}
\IEEEauthorblockA{\textsuperscript{1}\textit{Open Energy Transition, Königsallee 52, Bayreuth, 95448, Germany} \\\textsuperscript{2}\textit{Fraunhofer Research Institution for Energy Infrastructures and Geothermal Systems IEG, Cottbus, Germany}\\
\textsuperscript{3}\textit{Albert-Ludwigs Universität Freiburg, Faculty of Environment and Natural Resources, Freiburg im Breisgau, Germany} \\
\textsuperscript{4}\textit{Brandenburg University of Technology, Chair of Integrated Energy Infrastructures, Cottbus, Germany} \\ \textsuperscript{5}Technische Universität Berlin, Department of Digital Transformation in Energy Systems, Institute of Energy Technology\\
\textsuperscript{a}hazem.abdelkhalek@openenergytransition.org}
}


\maketitle

\begin{abstract}
Hydrogen is expected to play a key role in the energy transition. Analyses exploring the price of hydrogen usually calculate average or marginal production costs regardless of the time of delivery. A key factor that affects the price of hydrogen is the balancing costs, which we define as the expense of ensuring a steady schedule of hydrogen delivery. We explore the effect of delivering hydrogen to the export ports at different schedules, ranging from fully flexible to moderately stable with a daily and weekly buffer, to fully stable. We quantify the rise in hydrogen price with strict balancing constraint in three countries: Brazil, Morocco and Turkey, and three export volumes: 10, 50 and 200 TWh. The price difference between the flexible and stable schedules was found to reach a maximum of 36\% in Brazil, 47\% in Morocco and 18\% in Turkey across the different export volumes.
\end{abstract}
\vspace{0.4cm}

\section{Introduction}

Hydrogen is expected to play an important role in the global energy transition due to its versatility. Different factors affect the price of green hydrogen, including electricity prices, largely influenced by the renewable resource quality, equipment cost such as electrolysers and the interest rates, mirroring the risk perception of the project and country. One further key factor affecting the price of hydrogen is the balancing costs — which we define as the expense of ensuring a steady schedule of hydrogen delivery. Often overlooked in literature, balancing costs are particularly important as many practical applications which require a relatively stable supply of hydrogen and cannot accommodate large fluctuations.

Among these applications, which in most cases will have some requirements regarding time or steadiness of delivery, are hydrogen exports, which we specifically focus on in this paper. In this context, hydrogen balancing is important for pipeline exports as well as the synthesis of hydrogen-based carriers for exports by ships. Analyses exploring the cost of hydrogen usually calculate average or marginal production costs regardless of the time of delivery, which implicitly assumes full flexibility for the utilization or trading of hydrogen on the demand side. The additional cost for a steadier supply can significantly impact the overall cost of green hydrogen, given the variable nature of renewable electricity.

In isolated systems, balancing costs can straightforwardly be accounted for, as they are primarily tied to the costs of on-site hydrogen tanks or cavern storages. 

In contrast, integrated systems can balance hydrogen supply more flexibly by utilizing various storage technologies within the system and adjusting dispatch schedules as needed. However, this flexibility complicates calculating balancing costs, as it potentially distributes the efforts across multiple assets in the network, including battery storage and dispatchable power generators. 
In this study, we investigate several case studies to assess and quantify the cost of this balancing approach and the effect it has on different infrastructure components within integrated national systems.

\section{Literature Review}

\begin{figure*}[t!]
\centering
\begin{minipage}[b]{0.325\textwidth}
\centering
\includegraphics[width=\textwidth]{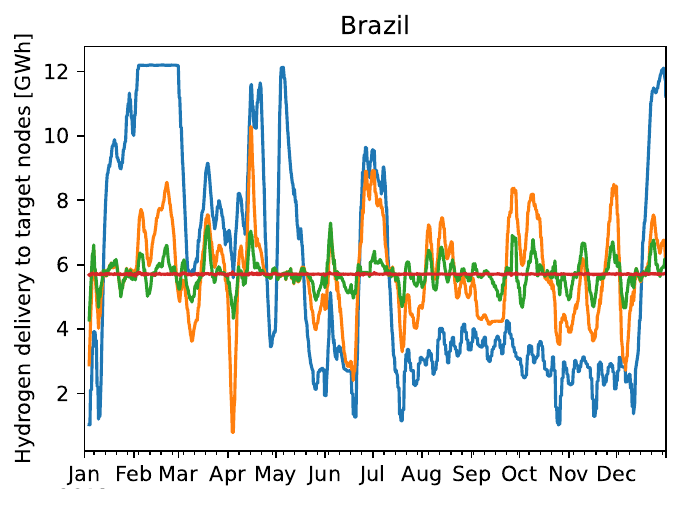}
\label{fig:}
\end{minipage}
\begin{minipage}[b]{0.325\textwidth}
\centering
\includegraphics[width=\textwidth]{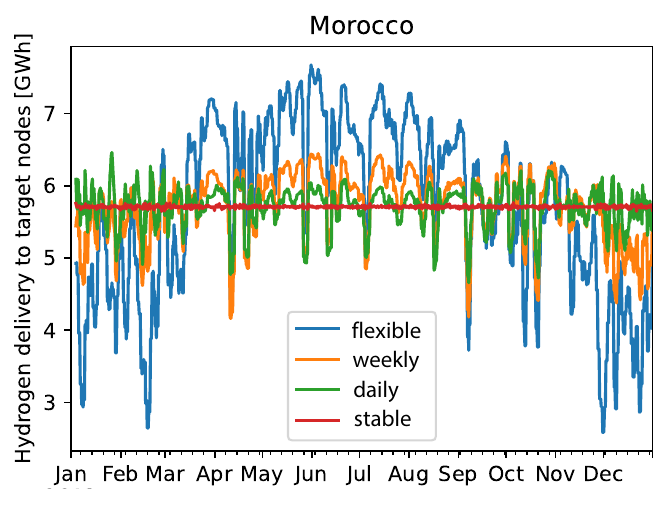}
\label{fig:}
\end{minipage}
%
\raisebox{0ex}{\begin{minipage}[b]{0.331\textwidth}
\centering
\includegraphics[width=\textwidth]{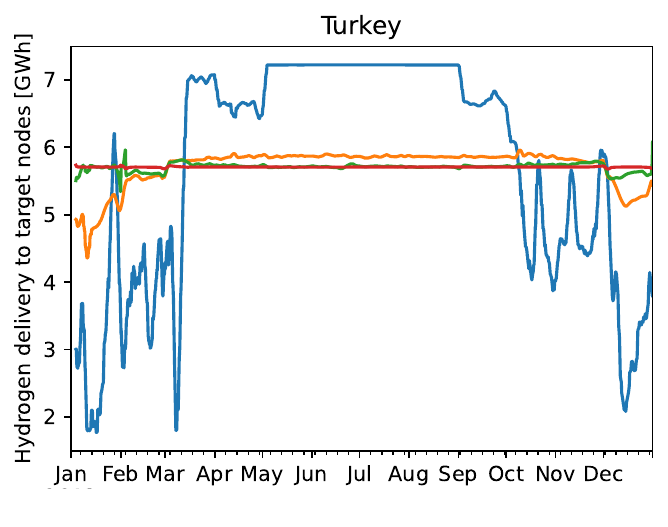}
\label{fig:}
\end{minipage}}
\caption{Optimal export delivery schedules under different flexibility constraints in Brazil, Morocco, and Turkey.}
\label{fig:delivery_schedule_3countries}
\end{figure*}

Several studies have assessed the techno-economic aspects of hydrogen export and quantified its cost. One example study, performed in Vietnam using HOMER Pro software, analyzed hydrogen storage and transmission for export to Japan and Korea. The study found that the levelized costs of hydrogen (LCOH) are USD 2.00/kg for liquid hydrogen, USD 1.86/kg for ammonia, and USD 1.72/kg for liquid organic hydrogen carriers. Liquid hydrogen ranks first in transmission cost but incurs the highest overall LCOH due to conversion and storage costs \cite{TA2024687}. 
Another study, conducted in Algeria, concludes the challenge for fossil-fuel-exporting countries, reporting that a 2050 renewable-only power system will not become a reality with out the development of renewable energy and supporting policies; approaches through which hydrogen can be incorporated into national decarbonization strategies—such as load shifting with electrolysis—are proven to boost efficiency \cite{MULLER2024240}. 

An independent study formulates an index-based analysis to estimate the global competitiveness of hydrogen exports, focusing on the United States, Australia, and China as the top countries that will likely dominate the projected 300~Mtons hydrogen market by the year 2050 \cite{HJEIJ20235843}. Another study investigates the economic feasibility of exporting renewable energy from island sites and shows that exporting electricity, at a yearly amount of 51 GWh for 42 million EUR, is more economically favorable than exporting hydrogen \cite{XU2023126750}. 

A more recent study \cite{MAKEPEACE20241183} performed a techno-economic assessment to show the economic viability of green hydrogen trade along 15 different international routes, concluding that 85\% of the global projected demand of 2050 will need to be traded internationally. Furthermore, it estimated the LCOH for countries with the best renewable energy conditions—such as Chile, Saudi Arabia, Australia, and the United States—to be in the range of 1.6–2.7~ USD/kg in 2030 and 1.4–2.3~USD/kg in 2040.

Globally focused studies complement this analysis by spatially explicit predictions about hydrogen cost prospects for 28 countries, find a total hydrogen potential of more than 1500~PWh/a, 79 PWh/a available at less than 2.30 EUR/kg until 2050. These insights highlight the superior role of regions with high solar resources, foremost in Africa and the Middle East, for sourcing low-cost hydrogen. It also hint at the strong need for ensuring secure water resources as well as balancing trade relationships toward better geostrategic supply security \cite{FRANZMANN202333062}.

Another study \cite{BRANDLE2021117481}, estimating the global supply cost of low-carbon hydrogen, concludes that steam-methane reforming (SMR) with carbon capture and storage (CCS) is the most cost-efficient in the medium term until 2030. It also concludes that renewable-based hydrogen can become competitive in the long run, between 2030 and 2050. The study assesses the LCOH based on the renewable energy source, finding the following LCOH ranges: For PV-based, around 2.0–2.7~USD/kg in 2030 and 1.0–1.8~USD/kg in 2050. For onshore wind, around 1.75–3.0~USD/kg in 2030 and 1.2–2.2~USD/kg in 2050.

While all these studies address the cost of hydrogen for export in different scenarios and from different countries across the world, they do not address the effect of the delivery schedule of hydrogen on these costs. In this study, we aim to address this research gap applying the methodology to three different countries and quantifying the effect of temporal flexibility of hydrogen delivery for export on the final price of hydrogen.

\vspace{0.2cm}
\section{Methodology}

The study utilizes PyPSA-Earth sector-coupled \cite{ABDELKHALEK2025125316} whose approach is a linear programming problem aimed at minimizing the total system cost by optimizing both expansion and dispatch of system components. 

We apply our methodology on three different countries – Brazil, Morocco, and Turkey– with distinct demand structures, renewable energy resources, existing infrastructure and hydrogen export ambitions. The export volumes considered are 10, 50, and 200~TWh in Brazil, Morocco, and Turkey. The scenario under study is a moderate scenario in terms of emission reduction and conservative in terms of hydrogen infrastructure rollout. It allows only ship exports and a short-term time horizon focusing on 2030. The energy systems of the three countries are modelled in high spatial (27-35 nodes per country) and temporal (3-hour timesteps) resolutions. The integrated modelling approach ensures the domestic demand of all sectors is satisfied before accounting for exports, utilizing output data from Sectoral LEnS \cite{muller2023national}. This is carried out in separate pre-export model runs where no export is required. The transmission grid of each country is sourced from OpenStreetMap \cite{openstreetmap_contributors_openstreetmap_2022}, and the existing power plants rely on the Powerplantmatching tool \cite{noauthor_powerplantmatching_2023} following the methodology of PyPSA-Earth \cite{parzen_pypsa-earth_2023}.
\begin{figure*}[ht!]
    \centering
    \includegraphics[width=\linewidth]{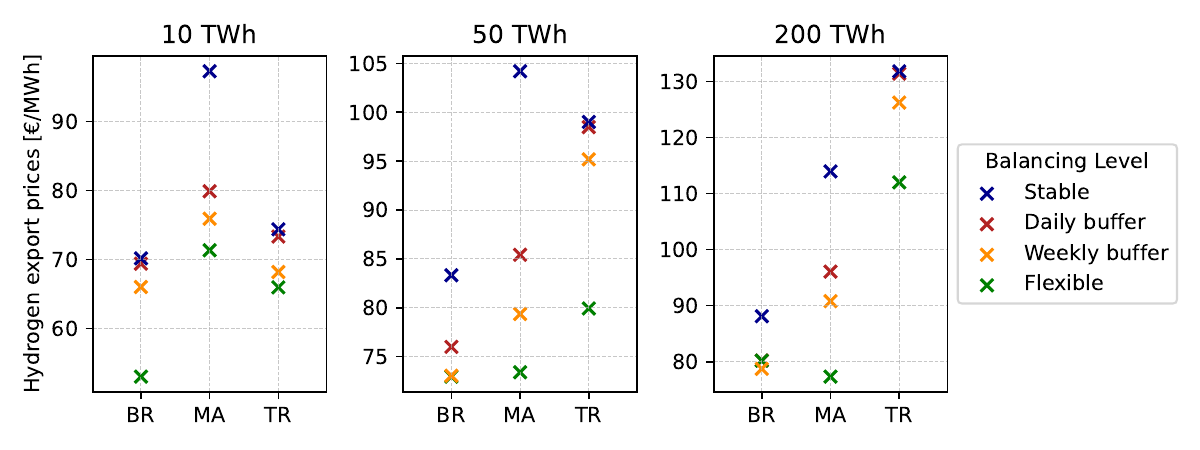}
    \caption{\textcolor{black}{Hydrogen export prices of 10, 50 and 200 TWh in Brazil, Morocco and Turkey under all flexibility constraints.}}
    \label{fig:balancing_all_costs}
\end{figure*}
The renewable potential calculator 2.0 of the Enertile model is used to calculate renewable energy potentials, their capital expenditure (CAPEX), and generation timeseries \cite{franke_assessing_2024}. The model uses ERA5 weather data of the year 2010 \cite{european_centre_for_medium-range_weather_forecasts_ecmwf_reanalysis_2020} with high geographical resolution (6.5 × 6.5 km$^2$ tiles), this tool assesses the potential of utility-scale PV, CSP, onshore wind, and offshore wind. The resulting renewable energy resource data is aggregated to the nodes of the countries assessed. GlobCover \cite{european_space_agency_globcover_2010} is used to identify feasible regions. The weighted average cost of capital (WACC) used in the study is different per country. For Brazil a WACC of 7.6\% is assumed, for Morocco 7.1\% and for Turkey 10\%. The allowed emission limit is 367, 54 and 313 Mtons of carbon dioxide equivalent in Brazil, Morocco and Turkey respectively. These values are in line with a linear trajectory to achieve an 80\% decrease compared to 2019 levels by 2050. Parameter list for key technologies is presented in table \ref{tab:parameters} in the appendix.

For exports, we assume four different levels of balancing ranging from total flexibility, to lenient, in which a daily or weekly buffer is assumed on the demand side, to strict, in which constant hydrogen delivery is enforced. The analysis is focused on the supply of green hydrogen aimed at exports and carried out for the defined export quantities for each country.  Equation \ref{eq:export_withbuffer} \cite{ABDELKHALEK2025125316} formulates the export feature of the model.

\begin{equation}
\begin{aligned}
Q = \sum_{t \in T} \dot{\epsilon}_{t} +  
\sum_{t \in T} \sum_{n \in N_{x}} \Bigg[
    \sum_{k \in K} \dot{C}_{n, H_2, k, t} 
    +\sum_{k \in K} \dot{E}_{n,H_2,k,t} 
   \\ + \sum_{l \in L^{in}_n} \dot{F}_{l, H_2, t} 
    - \sum_{l \in L^{out}_n} \dot{F}_{l,H_2, t} 
    -\sum_{s \in S} d_{n,H_2,s,t}
\Bigg]
\end{aligned}
\label{eq:export_withbuffer}
\end{equation}

Where $Q$ is the total hydrogen export volume, $n, N_x$ represent the node and set of export nodes in the system. In a similar manner $c, C$ denote carriers, $k, K$ represent technologies, and $t, T$ correspond to time-step and $l, L$ to the pipelines. $\dot{C}_{n,k,t}$ represents the dispatch of conversion technology $k$ at node $n$ and timestep $t$. Similarly,  $\dot{E}_{n,k,t}$ denotes the dispatch of storage technology and $\dot{F}$ represents the flow through pipelines. The demand at the node is denoted with $d_{n,H_2,s,t}$, where $s$ represents the different hydrogen consuming sectors. Most importantly, $\dot{\epsilon}_{t}$ is the maximum capacity of fictional hydrogen storage unit used to constrain on delivery schedule. 

\[
\dot{\epsilon}_{t} = \tau Q
\begin{cases} 
\tau = 1 & \text{for flexible delivery} \\
\tau = \frac{1}{52} & \text{for the weekly buffer} \\
\tau = \frac{1}{365} & \text{for the daily buffer} \\
\tau = \frac{1}{8760} & \text{for the stable supply} \\
\end{cases}
\]

To ensure the suitability for international green hydrogen markets, certain criteria are applied to the hydrogen production, mirroring the respective regulations. The main regulation is a monthly matching constraint to comply with additionality clauses. The restriction in the model enforces that the electricity used in hydrogen generation for export has to be matched with additional renewable electricity to the system. The spatial generation is not bound, assuming each country is one market zone. 
Equation \ref{eq:h2_policy} formulates the hydrogen policy constraint following the model's methodology and utilizing the pre-export runs \cite{ABDELKHALEK2025125316}.


\begin{equation}
\begin{aligned}
&\sum_{t \in M} \sum_{n \in N}  \left[\dot{C}_{n,EL,EY,t} - \dot{C}_{n,EL,EY,t}^{ref} \right] \quad\quad\quad\\
&\quad\quad\quad\quad\leq \sum_{t \in M}\sum_{n \in N}\sum_{k \in RES} \left[\dot{G}_{n,EL,k,t} - \dot{G}_{n,EL,k,t}^{ref} \right] \\[1.5ex]
\end{aligned}
\label{eq:h2_policy}
\end{equation}

where $\dot{C}_{n,EL,EY,t}$ represents the electricity $EL$ input to electrolyzer $EY$ at node $n$ and time $t$ and $\dot{G}_{n,EL, k, t}$ denotes the electricity dispatch of renewable energy power plants $k \in RES \subset K$ at node $n$ and time $t$. 
Additionally, $\dot{C}^{ref}$ and $\dot{G}^{ref}$ represent the same components but in the reference pre-export network with no hydrogen export required.
The two sides of the constraint are matched on a monthly basis, which is represented by the $M$ in the equation, where $M\subset T$ encompasses all time-steps of each calendar month.

The hydrogen prices considered in the study are the take-off prices at the export locations, meaning that any synthesis, liquefaction or shipping costs are not accounted for.



\vspace{0.2cm}
\section{Results}

\begin{figure*}[ht!]
    \centering
    \includegraphics[width=\linewidth]{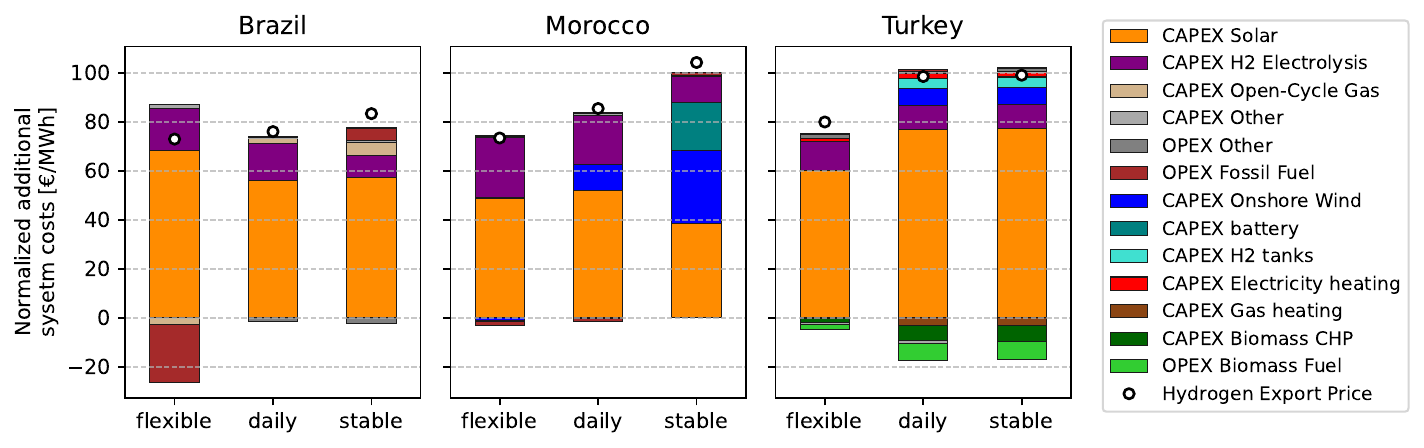}
    \caption{\textcolor{black}{Component-wise breakdown of normalized additional system costs triggered by hydrogen export in two scenarios.}}
    \label{fig:EEM_cost_comps_combined}
\end{figure*}

Figure \ref{fig:delivery_schedule_3countries} depicts the optimal hydrogen export schedule of 50 TWh of exports for each country under different flexibility constraints. The flexible schedules uniquely follow each country's renewable energy generation profile and system characteristics as much as allowed.
The flexible delivery schedule of Brazil, for example, highlights the influence of the rainy season, which typically occurs between December and April. Although in this study's weather year, heavy rainfall also occurred later in the year. Therefore, Brazil's flexible scenario captures pronounced hydropower peaks, which are utilized to generate low-cost hydrogen for export. Morocco's flexible schedule is affected by the solar pattern, as the country relies on solar PV for generating exports. Due to the smaller overall system size compared to the export volume investigated, its solar pattern follows the profile closely, even seasonally, as the generation fleet is sized specifically for export purposes. The electricity necessary to generate 50 TWh of hydrogen is approximately 71 TWh, which is 33\% higher than the total domestic consumption (54 TWh). In the case of Turkey, the heating season is the decisive factor for the favoured flexible schedule. The system favours delivering hydrogen for exports during the summer months, avoiding the load peaks necessary to satisfy the heat demand, thus creating a complementary load curve. Turkey has a large fleet of dispatchable generators, resulting in a higher chance to shift the renewable surplus over the months and increasing the electrolysis fleet's utilization, operating at full capacity for extended periods of time, as seen in figure \ref{fig:delivery_schedule_3countries}. Full results are presented in table \ref{tab:full_prices} in the appendix.

Figure \ref{fig:balancing_all_costs} shows how these delivery schedules translate into hydrogen price differences. The effects are large in small export volumes; at 10 TWh of exports in Brazil, for instance, the hydrogen marginal price in the weekly buffer case is 24.5\% higher than the flexible case, and rises to 32.4\% higher in the case of stable delivery. In Morocco, comparing the different schedules to the flexible case shows a widespread increase ranging between 6.4\% in the weekly buffer case and 36.4\% in the case of stable delivery. While the difference scheduling makes in both cases is significant, the underlying reason is, in fact, different. In the case of Brazil, the flexible delivery is the outlier as the system capitalizes on the hydro generation peaks occurring in the rainy season, which helps benefit the system. While in the case of Morocco, contrary to Brazil, the stable delivery is the outlier compared to the rest of the schedules due to the congestion of the transmission grid forcing the system to expand renewable capacities at suboptimal conditions. Turkey’s system benefits from inherently higher flexibility, resulting in less dramatic price variations ranging between 3.4\% and 12.8\%.

At 50 TWh of exports, the price difference tends to decrease for Brazil. The maximum deviation of 14.2\% is compared to the flexible schedule in the case of 50 TWh in Brazil. The difference between the flexible schedule and the more strict schedule diminishes as the hydro peaks are no longer enough to supply 50 TWh of hydrogen for export. In the case of Morocco and Turkey, the price difference gets larger, ranging between 8.1\% and 42\% in the case of Morocco, and 19.1\% and 23.9\% in the case of Turkey.
In the case of 200 TWh, the deviation in Brazil is limited to 9.9\%, while in Morocco the difference is spread between 17.4\% and 47.3\%, and in Turkey, with less deviation, remains between 12.7\% and 17.7\%.

Figure \ref{fig:EEM_cost_comps_combined} shows the normalized additional cost components to the system caused by export volumes of 50 TWh for all countries, including both capital and operational expenditure. This view allows for a better understanding of the factors driving the differences in export prices; it provides only an indication of the factors influencing marginal price formation and does not capture non-incremental costs, such as dynamic operational constraints like network congestion and scarcity rents. The comparison of the daily and constant delivery schedules to the flexible case gives further insight into the cause of the hydrogen price increase.
In Brazil, the system utilizes hydro peaks to generate hydrogen (see Figure \ref{fig:delivery_schedule_3countries}). Because this hydroelectricity was already generated in the no-export case (additionality constraint), it must generate renewable electricity to abide by the temporal matching. Hence, the system expands solar PV near the demand centers to cut down on gas turbines. With stricter schedules in Brazil, this is no longer the case, as the system hardly uses hydro for hydrogen production and thus does not necessarily need to offset fossil generation, as in the flexible case, to abide by the temporal constraint. This leads to lower prices in the flexible schedule compared to the stricter delivery schedules.

In Morocco, in the flexible case, the system relies solely on solar PV to generate hydrogen for export. As the delivery schedule becomes stricter, the system starts to rely more on onshore wind. This is consistent across both, the daily and stable scenarios, at 50 TWh. Even though onshore wind provides electricity at a higher cost than solar PV, it is more stable and aligns better geographically with the export ports, decreasing the need for long-distance power transport. Under the stable schedule, the inclusion of batteries is necessary to satisfy the required stable schedule of delivery, whether by balancing out the power generated at the electrolysis location and port or by balancing solar-based electricity at prime locations to better utilize the transmission grid. The behavior explained for the daily and stable schedules leads to higher electricity prices and consequently higher hydrogen costs.

In Turkey, the generation fleet is quite large and predominantly fossil-based. This is why the introduction of green hydrogen and the forced integration of renewables into the system impact many technologies. Excess renewable electricity from solar PV, expanded mostly for export purposes, is fed into the system through the same spillover effect observed in the previous sections. In Turkey, however, this electricity is primarily utilized in the heating sector: the system expands electricity-based heaters to absorb the excess, enabling a partial phase-out of both emission-producing technologies, such as gas boilers, and emission-reduction technologies, like biomass CHP CC. This behaviour remains consistent across all stricter schedules. Moreover, similar to Morocco, the system increasingly relies on more expensive onshore wind to maintain a steadier supply. This reliance, accompanied by the use of hydrogen tanks for storage and delivery, raises the overall hydrogen export price.


\section{Conclusions}

The time of delivery of hydrogen is a crucial factor in determining the competitiveness and economic viability of hydrogen exports. Allowing fully flexible schedules aligns production with each country’s dominant renewable resources and domestic demand, minimizing costs but resulting in highly irregular delivery patterns. This is evident in Brazil, where hydropower peaks during rainy seasons drive hydrogen production. This potentially leads to unfair pricing and increased costs in subsequent supply chain stages, in the form of necessary large storage facilities to accommodate large influxes of hydrogen during short periods of time. Enforcing stable delivery schedules at export points results in higher but more predictable hydrogen prices, as stricter supply constraints necessitate additional balancing mechanisms. 

In Brazil, price increases under stable schedules can reach 32\% for smaller export volumes but decrease to around 10\% for larger volumes due to solar resources near export points. In Morocco, where renewables are farther from ports and grid capacity is limited, price deviations range from 36\% for 10 TWh exports to 42\% under stable supply, while Turkey’s large dispatchable generation fleet and robust transmission network keep price variations between 17–20\%. 

The increased costs in stricter schedules stem from necessary system adaptations, such as additional generation and storage investments—backup capacity in Brazil, expanded wind and battery storage in Morocco, and added solar PV, wind, and hydrogen storage in Turkey. Since the study addresses hydrogen at the export ports before synthesis and shipping,the findings can extend beyond exports to any large-scale hydrogen-demanding facilities. These include domestic industries like ammonia and steel production, where quantifying flexibility of feed-in profiles can improve price assessments and inform decision-making. 

Ultimately, the study highlights that while flexible schedules are more cost-efficient by maximizing renewable utilization, they can lead to higher costs further down the supply chain which extends beyond the scope of this analysis. In contrast, stable schedules offer a more realistic and predictable price for exports.  When forecasting future hydrogen costs and prices, assuming full demand-side flexibility might lead to over-optimistic results, as in reality few, if any, off-takers of hydrogen do not have technical restrictions or preferences regarding the delivery profile. For comparing the results of studies in the field, it would seem advisable to a least explicitly state the assumptions in this regard. Calculating the costs for a standardized profile, like stable delivery, could be a better indicator of future hydrogen costs, as this diminishes the influence of certain cost-decreasing effects, that could not be realized in real-world applications.   

\section{Limitations}

Across all three systems studied, the specific delivery schedule—for example, for liquefaction or synthesis plants—is crucial in determining the true cost of hydrogen. Similarly, factors like fleet size, loading times, and intermediate storage sizing for exports via ships or throughput and compressors limitations for direct exports via pipelines play a significant role in shaping the optimal delivery schedule and significantly affecting costs.

One key limitation of this study is that we applied a single constraint in the aggregated hydrogen delivery for all export ports. This can lead to occasional spikes or dips in the single port's delivery schedules. Imposing port-specific constraints could yield a more accurate representation of the research question but it could also considerably increase computational complexity. This implies that the actual price can be higher than that shown in the results section. Another limitation is using country-specific interest rates to mirror each country's unique economic conditions. While this approach produces more realistic absolute results for each case, it complicates direct comparisons by introducing additional variables into the price calculations. Finally, in the study, the hydrogen policy applied is monthly temporal matching with additionality. A more strict temporal matching constraint, hourly, could better represent current regulations for 2030. For instance, in the Brazilian case, the emissions limit applied was not reached in the case of no exports, so under the stable-delivery constraint as exports required more stable power supply new fossil generators came online. Hence, "green" hydrogen exports in this one case have slightly increased the total emissions. Future work could address these limitations by adding location-specific constraints, exploring uniform financing assumptions for cross-country comparisons, and adopting stricter or more policy-aligned temporal matching rules to enhance the representation of green hydrogen modelling.








\clearpage



\section*{Appendix}\label{sec:appendix}

\begin{table}[h]
    \centering
    \caption{Comparison of hydrogen prices across different delivery schedules and countries}
    \begin{tabular}{llccc}
        \toprule
        Country & Delivery Schedule & \multicolumn{3}{c}{Price [€/MWh]} \\
        \cmidrule(lr){3-5}
                 &                   & 10 TWh  & 50 TWh  & 200 TWh  \\
        \midrule
        Brazil   & flexible          & 53.0    & 73.0    & 80.2    \\
                 & weekly            & 66.0    & 73.0    & 78.7    \\
                 & daily             & 69.4    & 76.0    & 80.2    \\
                 & stable            & 70.2    & 83.3    & 88.1    \\
        \midrule
        Morocco  & flexible          & 71.3    & 73.4    & 77.3    \\
                 & weekly            & 75.9    & 79.4    & 90.8    \\
                 & daily             & 79.9    & 85.4    & 96.1    \\
                 & stable            & 97.3    & 104.2   & 114.0   \\
        \midrule
        Turkey   & flexible          & 66.0    & 79.9    & 112.0   \\
                 & weekly            & 68.2    & 95.2    & 126.3   \\
                 & daily             & 73.3    & 98.5    & 131.4   \\
                 & stable            & 74.4    & 99.0    & 131.8   \\
        \bottomrule
    \end{tabular}
    \label{tab:full_prices}
\end{table}


\begin{table}[h!]
    \centering
    \caption{Energy Technology Parameters}
    \renewcommand{\arraystretch}{1.1}
    \begin{tabular}{|>{\centering\arraybackslash}p{2cm}|>{\centering\arraybackslash}p{1.5cm}|>{\centering\arraybackslash}p{1cm}|>{\centering\arraybackslash}p{2cm}|}
        \hline
        \textbf{Technology} & \textbf{Parameter} & \textbf{Value} & \textbf{Unit} \\ \hline
        \textbf{Solar PV} 
            & FOM         & 8.77    & EUR/kWel/year \\ \cline{2-4}
            & investment  & 438     & EUR/kWel \\ \cline{2-4}
            & lifetime    & 20      & years \\ \hline
        \textbf{Onshore wind} 
            & FOM         & 21      & EUR/kWel/year \\ \cline{2-4}
            & VOM         & 1.35    & EUR/MWh \\ \cline{2-4}
            & investment  & 1450    & EUR/kWel\_e \\ \cline{2-4}
            & lifetime    & 30      & years \\ \hline
        \textbf{Electrolysis} 
            & FOM         & 2       & \%/year \\ \cline{2-4}
            & efficiency  & 0.68    & per unit \\ \cline{2-4}
            & investment  & 450     & EUR/kWel \\ \cline{2-4}
            & lifetime    & 30      & years \\ \hline
        \textbf{Battery Storage} 
            & lifetime    & 10      & years \\ \cline{2-4}
            & investment  & 142     & EUR/kWh \\ \hline
        \textbf{Battery Inverter} 
            & FOM         & 0.34    & \%/year \\ \cline{2-4}
            & efficiency  & 0.96    & per unit \\ \cline{2-4}
            & investment  & 160     & EUR/kW \\ \cline{2-4}
            & lifetime    & 10      & years \\ \hline
        \textbf{Hydrogen Storage Tank} 
            & FOM         & 1.11    & \%/year \\ \cline{2-4}
            & investment  & 44.91   & EUR/kWh \\ \cline{2-4}
            & lifetime    & 30      & years \\ \hline
        \textbf{OCGT} 
            & FOM         & 1.78    & \%/year \\ \cline{2-4}
            & VOM         & 4.5     & EUR/MWh \\ \cline{2-4}
            & efficiency  & 0.4     & per unit \\ \cline{2-4}
            & investment  & 435.24  & EUR/kWel \\ \cline{2-4}
            & lifetime    & 25      & years \\ \hline
        \textbf{Biomass CHP} 
            & FOM             & 3.58    & \%/year \\ \cline{2-4}
            & VOM             & 2.11    & EUR/MWh\_e \\ \cline{2-4}
            & efficiency      & 0.3     & per unit \\ \cline{2-4}
            & efficiency-heat & 0.71    & per unit \\ \cline{2-4}
            & investment      & 3210.28 & EUR/kWel\_e \\ \cline{2-4}
            & lifetime        & 25      & years \\ \hline
        \textbf{Gas Boiler Steam} 
            & FOM         & 4.18    & \%/year \\ \cline{2-4}
            & VOM         & 1       & EUR/MWh \\ \cline{2-4}
            & efficiency  & 0.93    & per unit \\ \cline{2-4}
            & investment  & 45.45   & EUR/kWel \\ \cline{2-4}
            & lifetime    & 25      & years \\ \hline
    \end{tabular}
    \label{tab:parameters}
\end{table}

\end{document}